\documentclass[twocolumn,preprintnumbers,aps,showpacs,superscriptaddress,prl]{revtex4-1}

\usepackage{palatino}
\usepackage[latin1]{inputenc}
\usepackage{epsf}
\usepackage{amsmath,amssymb}
\usepackage{latexsym}
\usepackage{calc}
\usepackage{shadow}
\usepackage{epsfig}
\usepackage[pdftex, pdfstartview = {FitH}]{hyperref}
\usepackage{graphicx}

\newcommand{\nabnu}{\nabla_\nu}

\newcommand{\la}{\langle}
\newcommand{\ra}{\rangle}

\newcommand{\eps}{\epsilon}

\newcommand{\buX}{{\underline{\bf X}}}

\newcommand{\buA}{{\underline{\bf A}}}
\newcommand{\bur}{{\underline{\bf r}}}
\newcommand{\buR}{{\underline{\bf R}}}

\newcommand{\br}{{\bf r}}

\newcommand{\bX}{{\bf X}}
\newcommand{\bA}{{\bf A}}

\newcommand{\bR}{{\bf R}}

\newcommand{\hV}{{\hat{V}}}

\newcommand{\hT}{{\hat{T}}}

\newcommand{\hU}{{\hat{U}}}
\newcommand{\hH}{{\hat{H}}}
\newcommand{\pd}{\partial}

\newcommand{\ddt}{{\frac{\pd}{\pd t}}}

\newcommand{\mL}{{\mathcal L}}

\renewcommand\Im{\operatorname{Im}}

\begin{document}
\title{Is the molecular Berry phase an artifact of the Born-Oppenheimer approximation?}
\author{S. K. Min}
\affiliation{Max-Planck Institut f\"ur Mikrostrukturphysik, Weinberg 2,
D-06120 Halle, Germany}
\affiliation{Center for Superfunctional Materials, Department of Chemistry, 
Pohang University of Science and Technology, San 31, Hyojadong, Namgu, Pohang 
790-784, Korea}
\author{A. Abedi}
\affiliation{Max-Planck Institut f\"ur Mikrostrukturphysik, Weinberg 2,
D-06120 Halle, Germany}
\affiliation{European Theoretical Spectroscopy Facility (ETSF)}
\author{K. S. Kim}
\affiliation{Center for Superfunctional Materials, Department of Chemistry, 
Pohang University of Science and Technology, San 31, Hyojadong, Namgu, Pohang 
790-784, Korea}
\author{E.K.U. Gross}
\affiliation{Max-Planck Institut f\"ur Mikrostrukturphysik, Weinberg 2,
D-06120 Halle, Germany}
\affiliation{European Theoretical Spectroscopy Facility (ETSF)}

\date{\today}
\pacs{31.15.-p, 31.50.-x, 31.50.Gh}

\begin{abstract}
We demonstrate that the molecular Berry phase and the corresponding 
non-analyticity in the electronic Born-Oppenheimer wavefunction is, in 
general, not a true topological feature of the exact solution of the full 
electron-nuclear Schr\"odinger equation. For a numerically exactly solvable 
model we show that a non-analyticity, and the associated geometric phase, only 
appear in the limit of infinite nuclear mass, while a perfectly smooth 
behavior is found for any finite nuclear mass.
\end{abstract}

\maketitle
Geometric phases are ubiquitous in physics and chemistry, and some of the most 
fascinating phenomena in condensed matter science such as topological 
insulators~\cite{fu2007,hasan2010}, ferroelectrics~\cite{Resta1994,Xiao2010}, the 
Aharonov-Bohm effect~\cite{Aharonov1959} as well as conical intersections in molecules~\cite{Mead1992,Yarkony1996,Kendrick1997,juanes2005} are closely 
associated with Berry phases. 
Geometric phases may arise when the Hamiltonian of a system depends on a set 
of parameters $\buR$. 
In Berry's original definition~\cite{Berry1984}, this 
parameter set is allowed to change adiabatically, i.e. very slowly in time 
along a given path $C=\{\buR(t')|t'\in[t_0,t]\}$ such that the solution of the time-dependent Schr\"odinger equation
(TDSE),
\begin{align}
\label{eq:TDSEe}
i\ddt\Phi_n(\bur;t) = \hH^{BO}(\bur;\buR(t))\Phi_n(\bur;t),
\end{align}
by virtue of the adiabatic theorem, is given by
\begin{align}
\label{eq:adiabaticsolution}
\Phi_n\!(\bur;t)\!=\!e^{-i\!\int^t_{t_0}\!\!\eps^{BO}_n\!(\buR(t'))dt'}\!e^{-i\gamma_n(C)}\Phi^{BO}_n\!(\bur;\buR(t))
\end{align}
where
\begin{align}
\label{eq:BO}
\hH^{BO}(\bur;\buR)\Phi^{BO}_n(\bur;\buR)=\eps^{BO}_n(\buR)\Phi^{BO}_n(\bur;\buR).
\end{align}
In Eq.~(\ref{eq:adiabaticsolution}), the first exponent is a dynamical phase which appears naturally from the 
TDSE while the second exponent, $\gamma_n(C)$, is given in terms 
of the Berry connection,
\begin{align}
\label{eq:Berryconnection}
\bA^{BO}_{\nu,n}(\buR)&=\la\Phi^{BO}_n(\buR)|-i\nabnu\Phi^{BO}_n(\buR)\ra_\bur,
\end{align}
as a line integral along the path $C$
\begin{align}
\label{eq:Berryphase}
\gamma^{BO}_n(C) &= \int_C\sum_\nu\bA^{BO}_{\nu,n}(\buR)\cdot d\bR_\nu \nonumber\\
  &=\sum_\nu\int_{t_0}^t\! dt'\bA^{BO}_{\nu,n}(\buR(t'))\cdot\frac{d\bR_\nu(t')}{dt'}.
\end{align}
The notation $\la\cdots\ra_\bur$ indicates integration over $\bur$-space only.
All quantities in this Letter are in atomic units, 
and a bold value with underline represents a multi-dimensional vector, 
i.e. $\buX\equiv\{\bX_\nu|\nu=1,\cdots\}$.  
When $C$ becomes a closed loop, $\mL$, this line integral,
\begin{align}
\label{eq:BP}
\gamma^{BO}_n(\mL) &= \oint_{\mL}\sum_\nu\bA^{BO}_{\nu,n}(\buR)\cdot d\buR_\nu,
\end{align}
may give a non-vanishing value if the loop encloses a conical intersection (CI). 
The value of $\gamma_n^{BO}(\mL)$ does not depend on the shape of $\mL$ as long as the 
loop encloses the CI.
While the concept displayed in Eqs. (\ref{eq:TDSEe})-(\ref{eq:BP}) is 
completely general, i.e. may refer to any Hamiltonian that depends on a set of 
parameters, $\buR$, the specific case the notation in Eqs. 
(\ref{eq:TDSEe})-(\ref{eq:BP}) refers to is the molecular Berry phase 
appearing in the Born-Oppenheimer (BO) approximation. The latter is 
fundamental to all modern condensed matter theory. It derives from the fact 
that, in most cases, the nuclei move extremely slowly compared to the 
electrons. Hence, as a first step, it is reasonable to neglect the nuclear 
kinetic energy operator in the complete molecular Hamiltonian leading to the 
so-called BO Hamiltonian, 
\begin{align}
\label{eq:HBO}
\hH^{BO}(\bur;\buR)\!=\!\hT_e(\bur)\!+\!\hV_{ee}(\bur)\!+\!\hV_{en}(\bur,\buR)\!+\!\hV_{nn}(\buR).
\end{align}
Here $\hT_e$ is the electronic kinetic energy operator, $\hV_{ee}(\hV_{nn})$ 
is the repulsive electron-electron (nuclear-nuclear) interaction, and $\hV_{en}$ is the 
electron-nuclear Coulomb attraction.
The complete molecular wavefunction can then be 
approximated by the adiabatic ansatz
$\Psi^{BO}_{mol}(\bur,\buR)=\chi^{BO}_{kn}(\buR)\Phi^{BO}_n(\bur;\buR)$
where $\chi^{BO}_{kn}(\buR)$ satisfies the Schr\"odinger equation 
\begin{align}
\label{eq:BOchi}
\left[\sum_\nu\!\!\frac{(-i\nabnu\!+\!\bA^{BO}_{\nu,n})^2}{2M_\nu}\!+\!\tilde\eps^{BO}_n\!(\buR)\right]\!\!\chi^{BO}_{kn}\!(\buR)\!\!=\!\!E_{kn}\!\chi^{BO}_{kn}\!(\buR)
\end{align}
with the generalized BO potential energy surface,
\begin{align}
\label{eq:modBOPES}
\tilde\epsilon^{BO}_n&\!(\buR)\!\!=\la\Phi^{BO}_n(\buR)\!|\hH_{BO}(\buR)\!|\Phi^{BO}_n\!(\buR)\ra_\bur\!\nonumber\\
                                 &+\!\sum_\nu\!\!\frac{\la\nabnu\Phi^{BO}_n\!(\buR)|\nabnu\Phi^{BO}_n\!(\buR)\ra_\bur}{2M_\nu}\!-\!\sum_\nu\!\!\frac{\bA^{BO}_{\nu,n}\!(\buR)^2}{2M_\nu}.
\end{align}
Here $M_\nu$ is the mass of the $\nu$-th nucleus.  
After the seminal work of Mead and Truhlar~\cite{Mead1979}, a lot of attention has been 
devoted to the molecular Berry phase associated with the vector potential,
Eq.~(\ref{eq:Berryconnection})~\cite{Wilczek1989,Domcke2004book}. An essential aspect of the molecular 
geometric phase is that it always appears in the presence of some kind of
non-analyticity in the $\buR$-dependence of $\eps^{BO}_n(\buR)$ and 
$\Phi^{BO}_n(\bur;\buR)$. Similar to Cauchy's theorem in complex analysis, where 
a loop integral in the complex plane picks up a phase if the loop encloses a 
pole, the line integral in Eq.~(\ref{eq:BP}) may pick up a Berry phase if the 
loop encloses a CI of BO surfaces. Clearly, in the case of the 
molecular Berry phase, the parametric dependence of the Hamiltonian in Eq.~(\ref{eq:BO})
is the result of an approximation, the BO approximation. In the 
full molecular Hamiltonian, $\buR$ is a dynamical variable. The objective of 
this Letter is to investigate whether this very specific topological feature, 
this non-analyticity in the $\buR$-dependence of the wavefunction, 
only occurs within the BO approximation or whether it may 
survive as a feature of the full molecular wavefunction $\Psi_{mol}(\bur,\buR)$, 
i.e. as a true feature of nature. To investigate this question, we employ the 
recently derived framework of exact factorization of $\Psi_{mol}(\bur,\buR)$ 
~\cite{Gidopoulos2005,Hunter1975,Abedi2010,Abedi2012,Abedi2013}. This formulation lends itself as a natural framework 
because it leads to a Berry-type vector potential but without invoking the BO approximation.
 
Within this formulation, $\Psi^N_{mol}(\bur,\buR)$, the exact $N$-th 
eigenstate of the full molecular Schr\"odinger equation
$\hH(\bur,\buR)\Psi^N_{mol}(\bur,\buR) = E_N\Psi^N_{mol}(\bur,\buR)$,
can be factorized as a single product 
$\Psi^N_{mol}(\bur,\buR)\!\!=\!\!\chi_N(\buR)\Phi_N(\bur;\buR)$, where
$\Phi_N(\bur;\buR)$ satisfies the partial normalization condition, 
$\int\!\!d\bur |\Phi_N(\bur;\buR)|^2 = 1$. The equations which 
$\Phi_N(\bur;\buR)$ and $\chi_N(\buR)$ satisfy 
are
\begin{align}
\label{eq:exacteqe}
&\left[\!\hH^{BO}\!(\bur;\buR)\!+\!\hU_{en}\!(\bur,\buR)\!\right]\!\!\Phi_N\!(\bur;\buR)\!=\!\eps^{ex}_N\!(\buR)\Phi_N\!(\bur;\buR)\\
\label{eq:exacteqn}
&\left[\sum_\nu\!\!\frac{(-i\nabnu+\bA^{ex}_{\nu,N})^2}{2M_\nu}\!+\!\eps^{ex}_N(\buR)\right]\!\chi_N\!(\buR)\!\!=\!\!E_N\chi_N\!(\buR) 
\end{align}
where $\hU_{en}$ is 
an electron-nucleus coupling operator given by
\begin{align}
\label{eq:coup}
\hU_{en}(\bur,\buR) &= 
\sum_\nu\frac{1}{M_\nu}\left[\frac{(-i\nabnu-\bA^{ex}_{\nu,N})^2}{2}\right.\nonumber\\
&\left.+\left(\frac{-i\nabnu\chi_N}{\chi_N}\!+\!\bA^{ex}_{\nu,N}\right)\!\cdot\!(-i\nabnu\!-\!\bA^{ex}_{\nu,N})\right].
\end{align}
$\eps^{ex}_N(\buR)$ is defined as 
\begin{align}
\label{eq:exactPES}
\epsilon^{ex}_N&\!(\buR)\!\!=\la\Phi_N(\buR)|\hH_{BO}(\buR)|\Phi_N(\buR)\ra_\bur\!\nonumber\\
                                  &+\!\sum_\nu\!\!\frac{\la\nabnu\Phi_N(\buR)|\nabnu\Phi_N(\buR)\ra_\bur}{2M_\nu}\!-\!\sum_\nu\!\!\frac{\bA^{ex}_{\nu,N}(\buR)^2}{2M_\nu},
\end{align}
and 
\begin{align}
\label{eq:exactA}
\bA^{ex}_{\nu,N}(\buR)=\la\Phi_N(\buR)|-i\nabnu\Phi_N(\buR)\ra_\bur.  
\end{align}
Because $\eps^{ex}_N(\buR)$ and $\buA^{ex}_N(\buR)$ yield the exact many-body 
nuclear density, $|\chi_N(\buR)|^2\!=\!\int d\bur |\Psi^N_{mol}(\bur,\buR)|^2$, and the 
exact many-body nuclear current density, 
$1/M_\nu\!\left(\Im\!\left[\chi_N^*\nabnu\chi_N\right]\!+\!|\chi_N|^2\!\bA^{ex}_{\nu,N}\right)\!=\!1/M_\nu\!\left(\Im\!\!\int\!\! d\bur \Psi^{N*}_{mol}\nabnu\Psi^N_{mol}\right)$,
we can call $\eps^{ex}_N(\buR)$ and 
$\buA^{ex}_N(\buR)$ the exact scalar potential and the exact vector potential.
They are unique up to gauge transformations, 
$\chi_N(\buR)\rightarrow\chi_N(\buR) e^{iS(\buR)}$ and 
$\Phi_N(\buR)\rightarrow\Phi_N(\buR) e^{-iS(\buR)}$~\cite{Abedi2010,Abedi2012}. 

Both the exact molecular wavefunction 
$\Psi^N_{mol}(\bur,\buR)\!=\!\chi_N\!(\buR)\Phi_N(\bur;\buR)$ and the adiabatic 
approximation $\Psi^{BO}_{mol}\!(\bur,\buR)\!=\!\chi^{BO}_{kn}\!(\buR)\Phi^{BO}_n\!(\bur;\buR)$ 
are given in terms of a single product of a nuclear and an electronic 
wavefunction where the latter satisfies the partial normalization condition 
$\int\!|\Phi(\bur;\buR)|^2d\bur\!=\!1$ for each nuclear configuration $\buR$.
Both in the exact case and in the adiabatic approximation, the nuclear factor 
satisfies a standard Schr\"odinger equation (Eqs.~(\ref{eq:exacteqn}) and~(\ref{eq:BOchi}), respectively) 
with a vector potential (Eqs.~(\ref{eq:exactA}) and~(\ref{eq:Berryconnection}), respectively) and a 
scalar potential (Eqs.~(\ref{eq:exactPES}) and~(\ref{eq:modBOPES}), respectively) that formally follow 
the same expression. In particular, the vector potential is defined as a Berry 
connection in both cases. 
The only difference is that in the adiabatic approximation the Berry 
connection $\buA^{BO}_n(\buR)$, Eq.~(\ref{eq:Berryconnection}), and the BO 
potential energy surfaces $\tilde\eps^{BO}_n(\buR)$, Eq.~(\ref{eq:modBOPES}), are evaluated from the BO 
electronic wavefunction $\Phi^{BO}_n(\bur;\buR)$ and while the exact Berry 
connection $\buA^{ex}_N(\buR)$, Eq.~(\ref{eq:exactA}), and the exact potential 
energy surfaces $\eps^{ex}_N(\buR)$, Eq.~(\ref{eq:exactPES}), are evaluated with the exact electronic 
wavefunction coming from Eq.~(\ref{eq:exacteqe}).
In this sense $\buA^{ex}_N(\buR)$ represents a feature of the exact molecular 
wavefunction. Can this exact Berry connection have the fascinating topological 
structure that gives rise to a non-vanishing Berry phase? In other words, is 
the geometric phase found within the adiabatic approximation a true feature of 
nature, or is it merely an ``artifact'' of the BO approximation?

This is the question we are going to address in the following by studying a 
2-dimensional model system which, in the BO approximation, has CIs leading to 
a Berry phase, and which, at the same time, is simple enough to allow 
for a numerically exact solution.

The system consists of three ions and an electron. Two of the ions are fixed at 
$(\pm L/2,0)$, and the third ion as well as the electron 
are allowed to move in 2-dimensional space. Representing the positions of the moving ion 
and the electron as $\bR=(X,Y)$ and $\br=(x,y)$, respectively, the full Hamiltonian is
\begin{align}
\label{eq:fullH}
\hH(\br,\bR) =& -\frac{1}{2M}\nabla_\bR^2-\frac{1}{2}\nabla_\br^2+V_{en}(|\br-(\frac{L}{2},0)|)\nonumber\\
&+V_{en}(|\br-(-\frac{L}{2},0)|)+V_{en}(|\br-\bR|)\nonumber\\
&+V_{nn}(|\bR-(\frac{L}{2},0)|)+V_{nn}(|\bR-(-\frac{L}{2},0)|)\nonumber\\
&+V_{nn}(L)+(R/R_0)^4
\end{align}
where the first two terms are the kinetic energy operators for the moving ion and 
the electron, respectively, and the electron-nucleus interaction potential, 
$V_{en}(x)=-1/\sqrt{a+x^2}$, and the nucleus-nucleus interaction potential, 
$W_{nn}(x)=1/\sqrt{b+x^2}$, are represented as soft Coulomb potentials while 
the last term is added to make the system bound. Here, the origin is set as 
the center of the two fixed ions. 
We choose parameters $a$, $b$, $R_0$ and $L$ as 0.5, 10.0, 3.5 and 
$4\sqrt{3}/5$, respectively.  
Since the interaction potentials for the three ions are identical, we can expect that 
symmetry-induced degenerate states exist at equilateral positions 
$\bR^\pm_{eq}\!\!=\!\!(0,\pm Y_{eq})$ where $Y_{eq}\!\!=\!\!\sqrt{3}/2\cdot L=1.2$ with $D_{3h}$ point group 
symmetry. 

\begin{figure}
\begin{center}
\includegraphics[width=8.6cm]{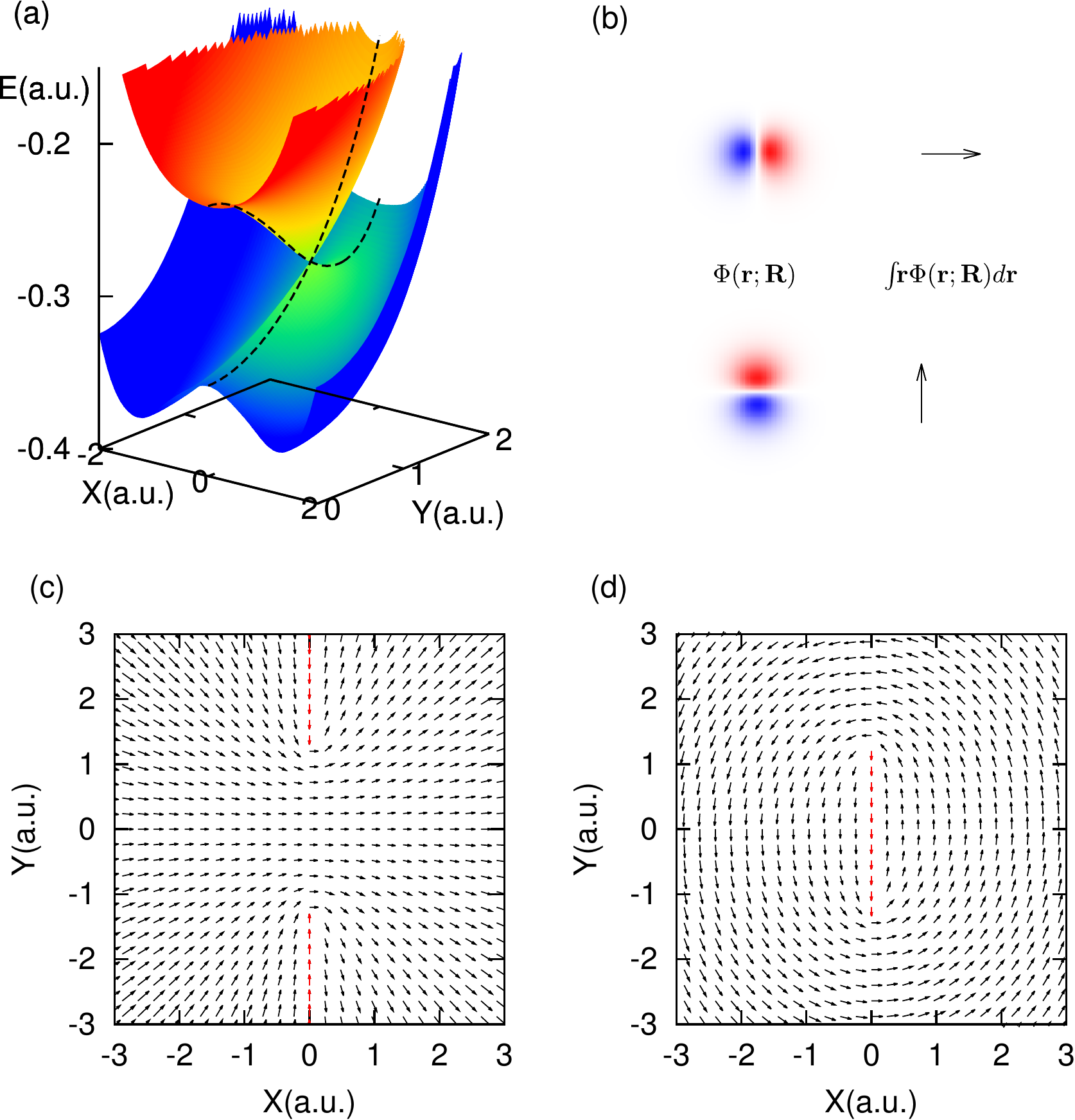}
\end{center}
\caption{(a) The first (blueish) and second (reddish) excited BO potential 
energy surfaces, (b) the vector field
representation for $p$-orbital-like wavefunctions at a certain $\bR$, and 
the BO electronic wavefunctions in the vector field representation for the first 
excited BO state (c) and the second excited BO state (d). The phase changes 
discontinuously across the line of red vectors ($L_1$ and $L_2$ in text).
}
\label{fig:H32P_BOstate}
\end{figure}

In FIG.~\ref{fig:H32P_BOstate}, we present the first and second excited BO surfaces, $\eps^{BO}_1(\bR)$ and 
 $\eps^{BO}_2(\bR)$, respectively, with the corresponding real-valued BO electronic 
wavefunctions $\Phi^{BO}_{1,2}(\br;\bR)$, which are numerical eigenstates of 
the BO Hamiltonian~\cite{note1}.
Indeed, we can confirm the degeneracy
between the energy levels $\eps^{BO}_1$ and $\eps^{BO}_2$ at $\bR^\pm_{eq}$ where the 
energy is -0.286.  
(Since the $s$-orbital-like ground BO electronic state is not related to CIs, 
we focus only on $\Phi^{BO}_{1,2}$.)
To visualize possible non-analyticities in the wavefunction 
$\Phi^{BO}_{1,2}(\br;\bR)$ with respect to $\bR$, we investigate the 
2-dimensional vector field
$\int\br\Phi(\br;\bR)d\br$ whose direction in space represents the 
``polarization of the wavefunction'': 
A $p$-orbital-like electronic wavefunction, $\Phi(\br;\bR)$,
can be represented as a vector pointing from the region of negative values of $\Phi(\br;\bR)$
to the region of positive values of $\Phi(\br;\bR)$ in $\br$-space as depicted in FIG.~\ref{fig:H32P_BOstate}(b).
The discontinuities of $\Phi(\br;\bR)$ appearing in $\bR$-space can then be seen as abrupt changes 
in the direction of the vectors.
We find that a discontinuous phase change
occurs across the lines $L_{1,2}$ for $\Phi^{BO}_{1,2}$, 
respectively, where $L_1\!\!=\!\!\{(X,Y)|X\!\!=\!\!0,|Y|\!\!>\!\!Y_{eq}\}$ and 
$L_2\!\!=\!\!\{(X,Y)|X\!\!=\!\!0,|Y|\!\!<\!\!Y_{eq}\}$    
(see red vectors in the lower panels of FIG.~\ref{fig:H32P_BOstate}). Along these lines
the sign of the $p$-orbital-like electronic wavefunctions, $\Phi^{BO}_{1,2}$, suddenly changes.
This leads to a non-zero Berry phase
($\gamma_n(\mL)=\pi$) if the closed path, $\mL$, crosses $L_1$ or $L_2$.

\begin{figure}
\begin{center}
\includegraphics[width=8.6cm]{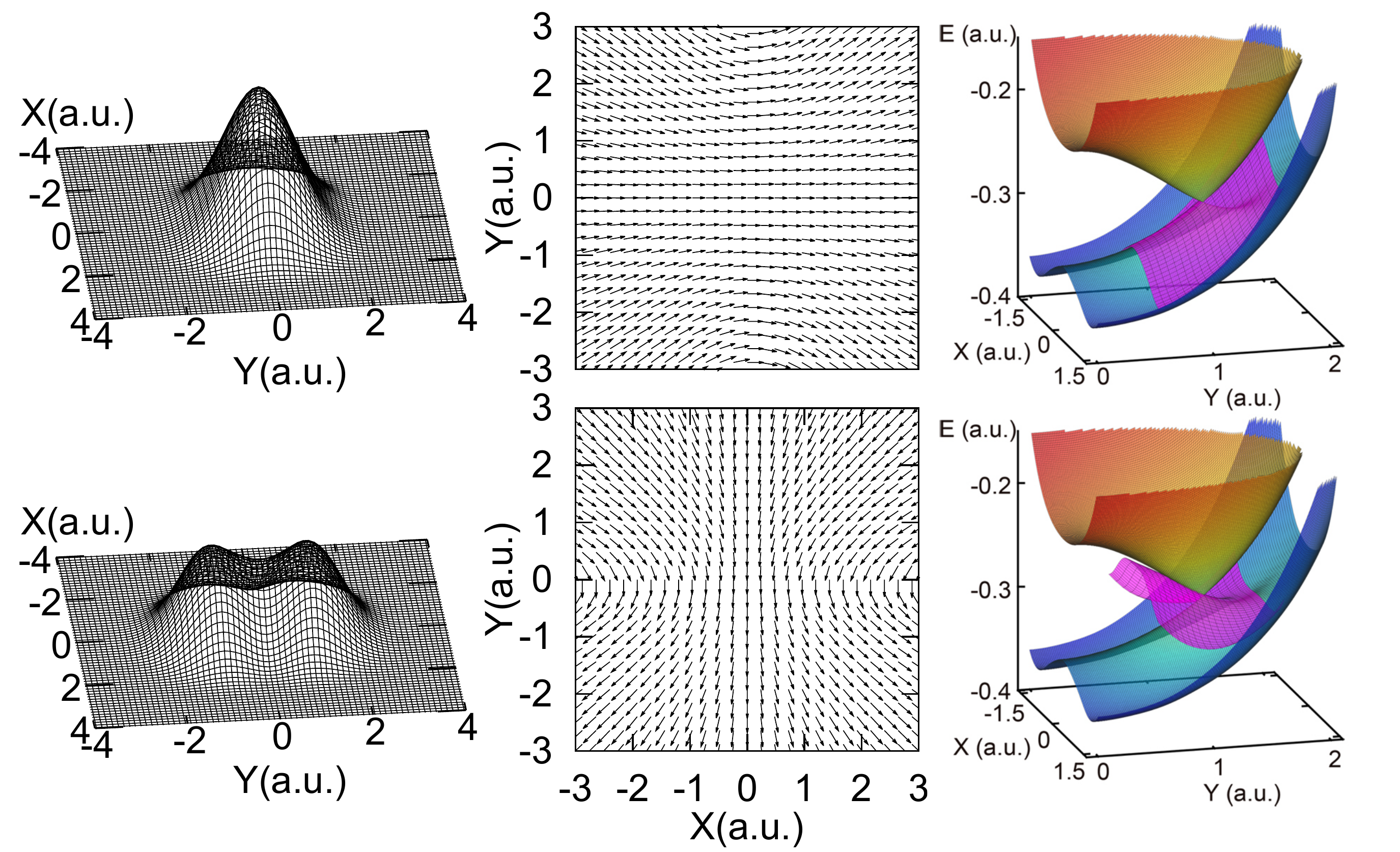}
\end{center}
\caption{The factorized nuclear densities (left), the corresponding 
electronic wavefunctions represented by the vector fields 
$\int\!\br\Phi_N(\br;\buR) d\br$ (middle) and the potential energy surfaces 
(right) for the selected full wavefunctions 
$\Psi^A_{mol}$ and $\Psi^B_{mol}$ (from top to bottom) with $M=10.0$.
The blueish and reddish surfaces are the first and second excited BO potential 
energy surfaces, 
respectively, while the pink surfaces are the exact potential energy surfaces near $\bR^+_{eq}$.  }
\label{fig:H32P_EXACT}
\end{figure}

In the exact decomposition framework, there is one potential energy surface for each 
exact eigenstate, $\Psi^N_{mol}(\br,\bR)$, of the full Hamiltonian, $\hH(\br,\bR)$. Here, we aim at 
investigating the behavior of the exact potential energy surfaces in the region at and around the points 
of CIs. To this end, we first calculate the eigenstates of the complete system up to a certain energy, 
well above the CIs involving the first and second excited BO potential energy surfaces. From the 
computed $\Psi^N_{mol}(\br,\bR)$, we calculate the exact nuclear wavefunction in a specific gauge as,
$\chi_N(\bR)=\sqrt{\int\!\!d\br|\Psi^N_{mol}\!(\br,\bR)|^2}$, and obtain the corresponding exact electronic 
wavefunction, $\Phi_N(\br;\buR)=\Psi^N_{mol}(\br,\bR)/\chi_N(\bR)$.  Then, for the subset of the exact electronic wavefunctions, 
$\Phi_N(\br;\buR)$, that exhibit $p$-orbital-like behavior similar to $\Phi^{BO}_1(\br;\bR)$ or $\Phi^{BO}_2(\br;\bR)$, 
we choose the energetically lowest two eigenstates, denoted as A and B, and 
calculate the exact potential energy surfaces $\eps^{ex}_{A}(\bR)$ and $\eps^{ex}_B(\bR)$ from 
Eq.~(\ref{eq:exactPES}). 
In FIG.~\ref{fig:H32P_EXACT}, we have plotted $|\chi_N(\bR)|^2$, 
$\int\br\Phi_N(\br;\buR)d\br$ and $\eps^{ex}_N(\bR)$ ($N\!=\!A,B$) for a 
nuclear mass of
$M\!\!=\!\!10.0$. 
The eigenenergies of $\Psi^{A,B}_{mol}$ are -0.282 and -0.201, 
respectively. As it is seen in FIG.~\ref{fig:H32P_EXACT}, for these 
exact eigenstates, $\chi_N(\bR^\pm_{eq})\ne 0$, and $\Phi_N(\br;\buR)$ do not show any abrupt phase 
changes or singularities. Therefore, $\eps^{ex}_A(\bR)$ and $\eps^{ex}_B(\bR)$ 
show a smooth ``diabatic'' 
form connecting $\eps^{BO}_1$ and $\eps^{BO}_2$ continuously along the $Y$-axis 
through the points where, in the BO case, are CIs. Consequently, the exact geometric phase 
$\gamma^{ex}_N(\mL)=\oint_\mL\bA^{ex}_N(\bR')\cdot d\bR'$ is zero
since there is no singular point in $\Phi_N(\br;\buR)$ in this case.  

\begin{figure}
\begin{center}
\includegraphics[width=8.6cm]{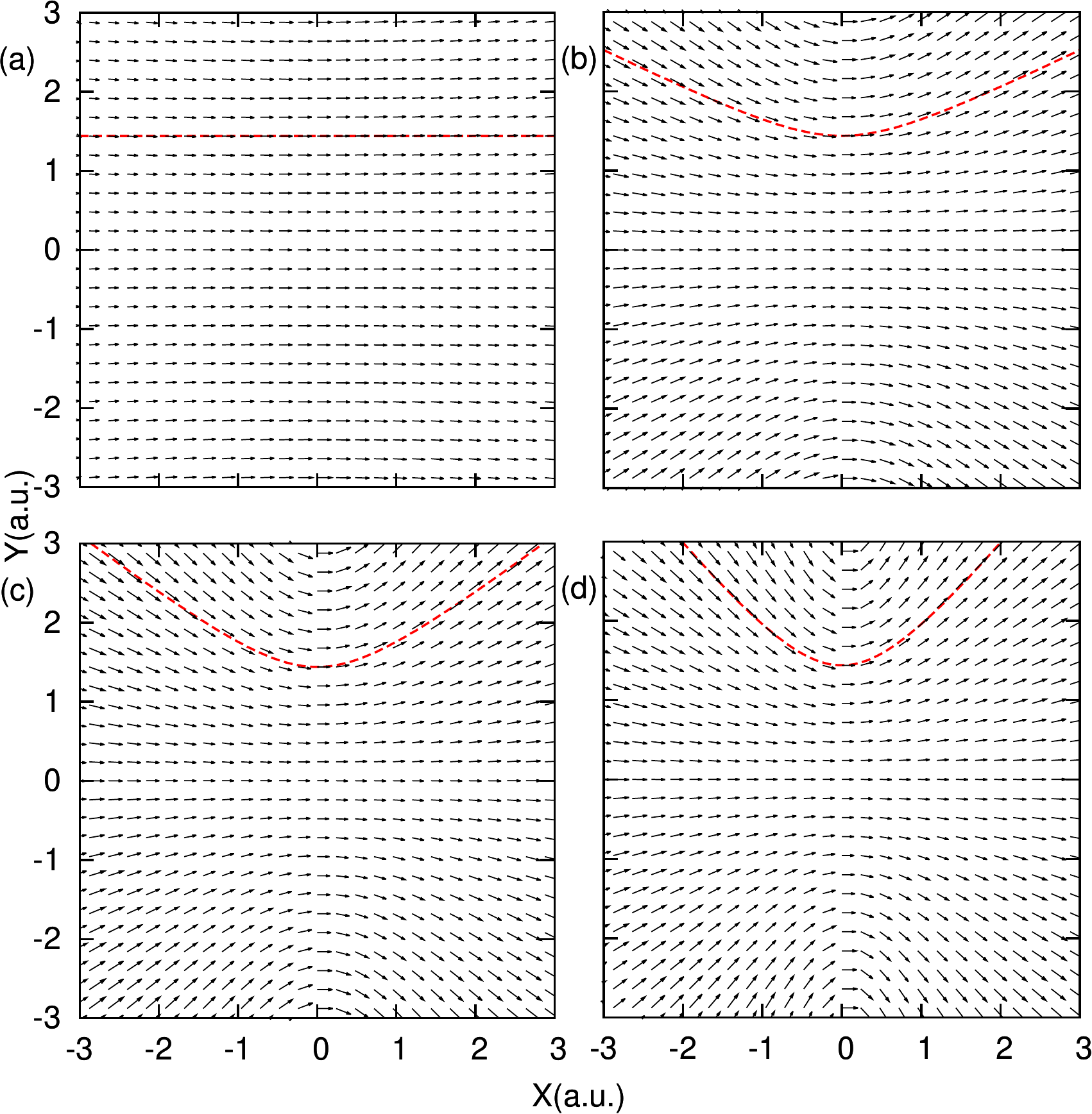}
\end{center}
\caption{The vector $\int \br\Phi_A(\br;\bR)d\br$ is plotted for various ionic 
masses, $M$. The values of $M$ for th panels (a), (b), (c) and (d) are 1.0, 10.0, 20.0 and 50.0, 
respectively. The dashed red line indicates, as guide for the eye, the change of curvature as the ionic 
mass increases.   
}
\label{fig:H32P_mass_dep_PHI}
\end{figure}

\begin{figure}
\begin{center}
\includegraphics[width=8.6cm]{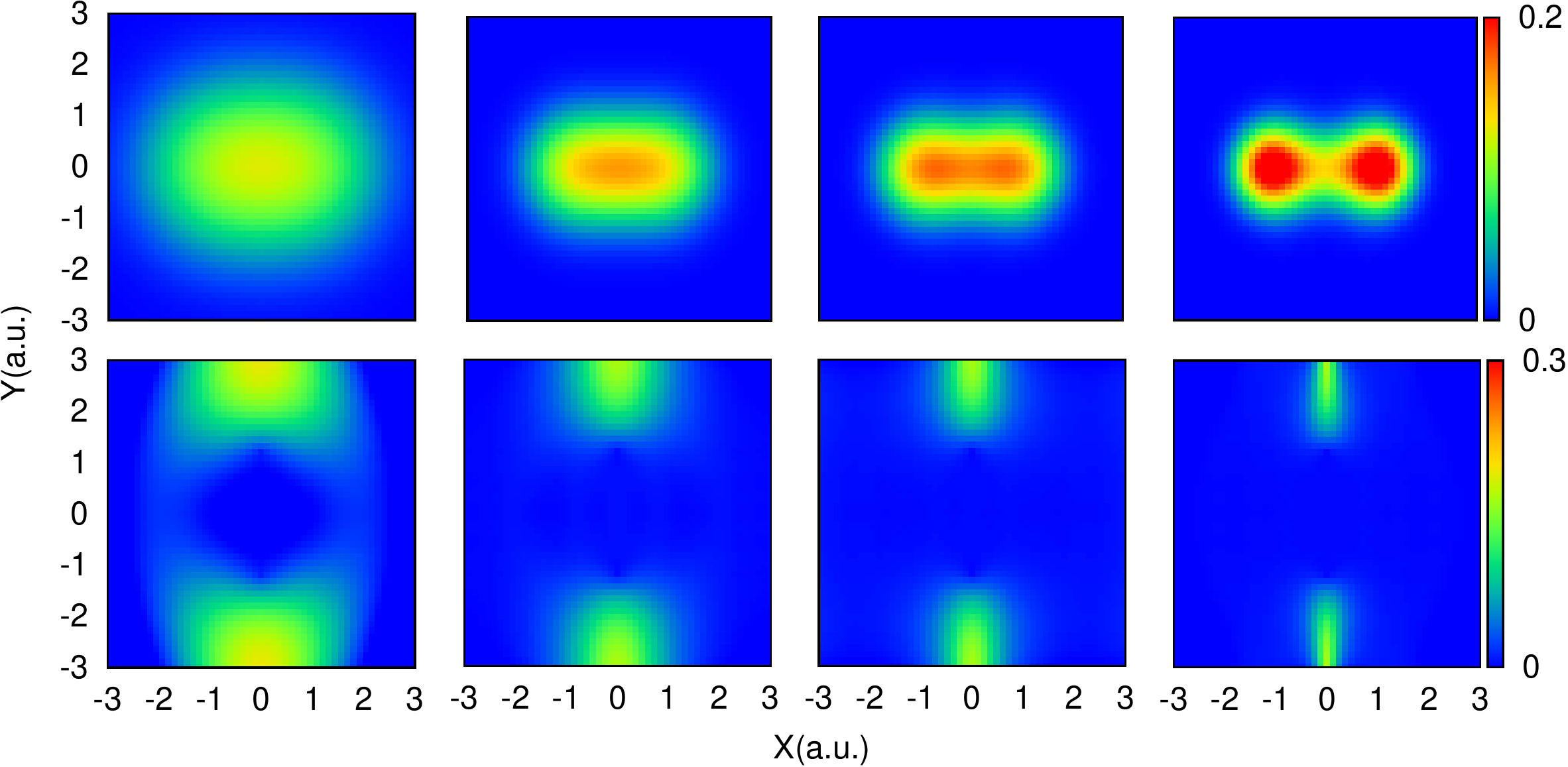}
\end{center}
\caption{The factorized nuclear densities (first low), and the difference 
between the exact potential energy surface and 
the 1st excited BO potential energy surface ($\epsilon^{ex}_A-\epsilon^{BO}_1$) (second row)
for various nuclear masses (M=1.0,10.0,20.0, and 50.0 from left to right).
}
\label{fig:H32P_mass_dep}
\end{figure}

In the following we investigate
how $\Phi_A(\br;\buR)$, $|\chi_A(\bR)|^2$ 
and $\eps^{ex}_A(\bR)$ evolve as $M$ increases ($M$=1,10,20, and 50) to reach the adiabatic 
limit ($M\rightarrow\infty$) which is accompanied by a Berry phase. 
In FIG.~\ref{fig:H32P_mass_dep_PHI}, 
we show how the exact electronic wavefunction $\Phi_{A}(\bur;\buR)$ transforms 
into $\Phi^{BO}_1$ with increasing nuclear mass.  
For $M\!=\!1$, a set of 
vectors representing the vector field shows a mainstream simply from left to right. As $M$ increases, 
however, the mainstream begins to show parabolic behavior, and the curvature  
of the parabola increases gradually. Compared to $\Phi^{BO}_1$ in FIG.~\ref{fig:H32P_BOstate}, 
we can interpret the discontinuity along $L_1$ as coming
from the infinite-curvature limit due to the limit $M\rightarrow\infty$.
In FIG. \ref{fig:H32P_mass_dep}, we also show $|\chi_A|^2$ and 
$|\eps^{ex}_A-\eps^{BO}_1|$ for various $M$'s.
As $M$ increases, $|\chi_A|^2$ gets localized on the double-minima of $\eps^{ex}_A$ 
and also gets narrower, showing two distinctive humps.
For $\eps^{ex}_A$, the green region around $L_1$ shrinks as $M$ increases, 
which means $\eps^{ex}_A$ gets closer to $\eps^{BO}_1$, but 
maintaining the diabatic behavior along the $Y$-axis. This enables us to deduce that  
$\eps^{ex}_A$ in the limit $M\!\!\rightarrow\!\!\infty$ lies
on top of the BO potential energy surface $\eps^{BO}_1$ except for the line $L_1$.
Since the actual nuclear mass in the real world is finite, there is no discontinuity of 
the electronic wavefunction implying that the exact geometric phase is zero.

To summarize, we have investigated whether the specific non-analyticity in 
$\Phi^{BO}_n(\bur;\buR)$ that leads to a non-trivial geometric phase in the BO 
approximation is a true topological feature of the full electron-nuclear 
wavefunction. To shed light on this question, we have studied a numerically 
exactly solvable model system in 2 dimensions that exhibits non-trivial Berry 
phases in the BO limit. Employing the exact factorization of the full 
molecular 
wavefunction~\cite{Gidopoulos2005,Hunter1975,Abedi2010,Abedi2012,Abedi2013}  
we identify and calculate the exact electronic wavefunctions $\Phi_N(\bur;\buR)$ 
which, in the limit of infinite nuclear mass $M$, reduce to the BO electronic 
wavefunctions $\Phi^{BO}_n(\bur;\buR)$. We find that the exact electronic 
wavefunctions $\Phi_N(\bur;\buR)$ are perfectly smooth for any finite value of 
the nuclear mass. Consequently the geometric phase associated with the vector 
potential 
$\bA^{ex}_{\nu,N}(\buR)=-i\int\Phi^*_N(\bur;\buR)\nabnu\Phi_N(\bur;\buR)d\bur$
vanishes. Only in the limit $M\rightarrow\infty$ (the BO limit) a 
discontinuous phase change appears which leads to a non-trivial Berry phase. 
In this sense, the molecular Berry phase can be viewed as an artifact of the 
BO approximation. The specific topological feature, the non-analyticity of 
the BO electronic wavefunction leading to the BO geometric phase is not a 
feature of the true molecular wavefunction. Whether or not this statement 
holds true for all molecules and solids is currently not known. One easily 
verifies that nodes at specific nuclear configurations $\buR_0$ in the exact 
molecular wavefunction, 
$\Psi^N_{mol}(\bur,\buR)$, directly lead to 
singularities in the exact vector potential $\buA^{ex}_N(\buR)$. Whether such 
singularities can produce non-trivial Berry phases remains the subject of 
future research.

We acknowledte partial support from the Deutsche Forschungsgemeinschaft (SFB 762), the 
European Commission (FP7-NMP-CRONOS), NRF (National Honor Scientist Program: 
2010-0020414) and KISTI (KSC-2011-G3-02)

\addcontentsline{toc}{section}{References}
\bibliography{./SKM140202}

 \newcommand{\noop}[1]{} \bibliographystyle{aipnum4-1}
\begin{thebibliography}{19}%
\makeatletter
\providecommand \@ifxundefined [1]{%
 \@ifx{#1\undefined}
}%
\providecommand \@ifnum [1]{%
 \ifnum #1\expandafter \@firstoftwo
 \else \expandafter \@secondoftwo
 \fi
}%
\providecommand \@ifx [1]{%
 \ifx #1\expandafter \@firstoftwo
 \else \expandafter \@secondoftwo
 \fi
}%
\providecommand \natexlab [1]{#1}%
\providecommand \enquote  [1]{``#1''}%
\providecommand \bibnamefont  [1]{#1}%
\providecommand \bibfnamefont [1]{#1}%
\providecommand \citenamefont [1]{#1}%
\providecommand \href@noop [0]{\@secondoftwo}%
\providecommand \href [0]{\begingroup \@sanitize@url \@href}%
\providecommand \@href[1]{\@@startlink{#1}\@@href}%
\providecommand \@@href[1]{\endgroup#1\@@endlink}%
\providecommand \@sanitize@url [0]{\catcode `\\12\catcode `\$12\catcode
  `\&12\catcode `\#12\catcode `\^12\catcode `\_12\catcode `\%12\relax}%
\providecommand \@@startlink[1]{}%
\providecommand \@@endlink[0]{}%
\providecommand \url  [0]{\begingroup\@sanitize@url \@url }%
\providecommand \@url [1]{\endgroup\@href {#1}{\urlprefix }}%
\providecommand \urlprefix  [0]{URL }%
\providecommand \Eprint [0]{\href }%
\providecommand \doibase [0]{http://dx.doi.org/}%
\providecommand \selectlanguage [0]{\@gobble}%
\providecommand \bibinfo  [0]{\@secondoftwo}%
\providecommand \bibfield  [0]{\@secondoftwo}%
\providecommand \translation [1]{[#1]}%
\providecommand \BibitemOpen [0]{}%
\providecommand \bibitemStop [0]{}%
\providecommand \bibitemNoStop [0]{.\EOS\space}%
\providecommand \EOS [0]{\spacefactor3000\relax}%
\providecommand \BibitemShut  [1]{\csname bibitem#1\endcsname}%
\let\auto@bib@innerbib\@empty
\bibitem [{\citenamefont {Fu}\ \emph {et~al.}(2007)\citenamefont {Fu},
  \citenamefont {Kane},\ and\ \citenamefont {Mele}}]{fu2007}%
  \BibitemOpen
  \bibfield  {author} {\bibinfo {author} {\bibfnamefont {L.}~\bibnamefont
  {Fu}}, \bibinfo {author} {\bibfnamefont {C.~L.}\ \bibnamefont {Kane}}, \ and\
  \bibinfo {author} {\bibfnamefont {E.~J.}\ \bibnamefont {Mele}},\ }\href@noop
  {} {\bibfield  {journal} {\bibinfo  {journal} {Phys. Rev. Lett.}\ }\textbf
  {\bibinfo {volume} {98}},\ \bibinfo {pages} {106803} (\bibinfo {year}
  {2007})}\BibitemShut {NoStop}%
\bibitem [{\citenamefont {Hasan}\ and\ \citenamefont {Kane}(2010)}]{hasan2010}%
  \BibitemOpen
  \bibfield  {author} {\bibinfo {author} {\bibfnamefont {M.~Z.}\ \bibnamefont
  {Hasan}}\ and\ \bibinfo {author} {\bibfnamefont {C.~L.}\ \bibnamefont
  {Kane}},\ }\href@noop {} {\bibfield  {journal} {\bibinfo  {journal} {Rev.
  Mod. Phys.}\ }\textbf {\bibinfo {volume} {82}},\ \bibinfo {pages} {3045}
  (\bibinfo {year} {2010})}\BibitemShut {NoStop}%
\bibitem [{\citenamefont {Resta}(1994)}]{Resta1994}%
  \BibitemOpen
  \bibfield  {author} {\bibinfo {author} {\bibfnamefont {R.}~\bibnamefont
  {Resta}},\ }\href {\doibase 10.1103/RevModPhys.66.899} {\bibfield  {journal}
  {\bibinfo  {journal} {Rev. Mod. Phys.}\ }\textbf {\bibinfo {volume} {66}},\
  \bibinfo {pages} {899} (\bibinfo {year} {1994})}\BibitemShut {NoStop}%
\bibitem [{\citenamefont {Xiao}\ \emph {et~al.}(2010)\citenamefont {Xiao},
  \citenamefont {Chang},\ and\ \citenamefont {Niu}}]{Xiao2010}%
  \BibitemOpen
  \bibfield  {author} {\bibinfo {author} {\bibfnamefont {D.}~\bibnamefont
  {Xiao}}, \bibinfo {author} {\bibfnamefont {M.-C.}\ \bibnamefont {Chang}}, \
  and\ \bibinfo {author} {\bibfnamefont {Q.}~\bibnamefont {Niu}},\ }\href
  {\doibase 10.1103/RevModPhys.82.1959} {\bibfield  {journal} {\bibinfo
  {journal} {Rev. Mod. Phys.}\ }\textbf {\bibinfo {volume} {82}},\ \bibinfo
  {pages} {1959} (\bibinfo {year} {2010})}\BibitemShut {NoStop}%
\bibitem [{\citenamefont {Aharonov}\ and\ \citenamefont
  {Bohm}(1959)}]{Aharonov1959}%
  \BibitemOpen
  \bibfield  {author} {\bibinfo {author} {\bibfnamefont {Y.}~\bibnamefont
  {Aharonov}}\ and\ \bibinfo {author} {\bibfnamefont {D.}~\bibnamefont
  {Bohm}},\ }\href {\doibase 10.1103/PhysRev.115.485} {\bibfield  {journal}
  {\bibinfo  {journal} {Phys. Rev.}\ }\textbf {\bibinfo {volume} {115}},\
  \bibinfo {pages} {485} (\bibinfo {year} {1959})}\BibitemShut {NoStop}%
\bibitem [{\citenamefont {Mead}(1992)}]{Mead1992}%
  \BibitemOpen
  \bibfield  {author} {\bibinfo {author} {\bibfnamefont {C.~A.}\ \bibnamefont
  {Mead}},\ }\href {\doibase 10.1103/RevModPhys.64.51} {\bibfield  {journal}
  {\bibinfo  {journal} {Rev. Mod. Phys.}\ }\textbf {\bibinfo {volume} {64}},\
  \bibinfo {pages} {51} (\bibinfo {year} {1992})}\BibitemShut {NoStop}%
\bibitem [{\citenamefont {Yarkony}(1996)}]{Yarkony1996}%
  \BibitemOpen
  \bibfield  {author} {\bibinfo {author} {\bibfnamefont {D.~R.}\ \bibnamefont
  {Yarkony}},\ }\href {\doibase 10.1103/RevModPhys.68.985} {\bibfield
  {journal} {\bibinfo  {journal} {Rev. Mod. Phys.}\ }\textbf {\bibinfo {volume}
  {68}},\ \bibinfo {pages} {985} (\bibinfo {year} {1996})}\BibitemShut
  {NoStop}%
\bibitem [{\citenamefont {Kendrick}(1997)}]{Kendrick1997}%
  \BibitemOpen
  \bibfield  {author} {\bibinfo {author} {\bibfnamefont {B.}~\bibnamefont
  {Kendrick}},\ }\href@noop {} {\bibfield  {journal} {\bibinfo  {journal}
  {Phys. Rev. Lett.}\ }\textbf {\bibinfo {volume} {79}},\ \bibinfo {pages}
  {2431} (\bibinfo {year} {1997})}\BibitemShut {NoStop}%
\bibitem [{\citenamefont {Juanes-Marcos}\ \emph {et~al.}(2005)\citenamefont
  {Juanes-Marcos}, \citenamefont {Althorpe},\ and\ \citenamefont
  {Wrede}}]{juanes2005}%
  \BibitemOpen
  \bibfield  {author} {\bibinfo {author} {\bibfnamefont {J.~C.}\ \bibnamefont
  {Juanes-Marcos}}, \bibinfo {author} {\bibfnamefont {S.~C.}\ \bibnamefont
  {Althorpe}}, \ and\ \bibinfo {author} {\bibfnamefont {E.}~\bibnamefont
  {Wrede}},\ }\href@noop {} {\bibfield  {journal} {\bibinfo  {journal}
  {Science}\ }\textbf {\bibinfo {volume} {309}},\ \bibinfo {pages} {1227}
  (\bibinfo {year} {2005})}\BibitemShut {NoStop}%
\bibitem [{\citenamefont {Berry}(1984)}]{Berry1984}%
  \BibitemOpen
  \bibfield  {author} {\bibinfo {author} {\bibfnamefont {M.~V.}\ \bibnamefont
  {Berry}},\ }\href@noop {} {\bibfield  {journal} {\bibinfo  {journal} {Proc.
  R. Soc. London, Ser. A}\ }\textbf {\bibinfo {volume} {392}},\ \bibinfo
  {pages} {45} (\bibinfo {year} {1984})}\BibitemShut {NoStop}%
\bibitem [{\citenamefont {Mead}\ and\ \citenamefont
  {Truhlar}(1979)}]{Mead1979}%
  \BibitemOpen
  \bibfield  {author} {\bibinfo {author} {\bibfnamefont {C.}~\bibnamefont
  {Mead}}\ and\ \bibinfo {author} {\bibfnamefont {D.}~\bibnamefont {Truhlar}},\
  }\href@noop {} {\bibfield  {journal} {\bibinfo  {journal} {J. Chem. Phys.}\
  }\textbf {\bibinfo {volume} {70}},\ \bibinfo {pages} {2284} (\bibinfo {year}
  {1979})}\BibitemShut {NoStop}%
\bibitem [{\citenamefont {Wilczek}\ and\ \citenamefont
  {Shapere}(1989)}]{Wilczek1989}%
  \BibitemOpen
  \bibfield  {author} {\bibinfo {author} {\bibfnamefont {F.}~\bibnamefont
  {Wilczek}}\ and\ \bibinfo {author} {\bibfnamefont {A.}~\bibnamefont
  {Shapere}},\ }\href@noop {} {\emph {\bibinfo {title} {Geometric phases in
  physics}}},\ Vol.~\bibinfo {volume} {5}\ (\bibinfo  {publisher} {World
  Scientific},\ \bibinfo {year} {1989})\BibitemShut {NoStop}%
\bibitem [{\citenamefont {Domcke}\ \emph {et~al.}(2004)\citenamefont {Domcke},
  \citenamefont {Yarkony},\ and\ \citenamefont {K{\"o}ppel}}]{Domcke2004book}%
  \BibitemOpen
  \bibfield  {author} {\bibinfo {author} {\bibfnamefont {W.}~\bibnamefont
  {Domcke}}, \bibinfo {author} {\bibfnamefont {D.}~\bibnamefont {Yarkony}}, \
  and\ \bibinfo {author} {\bibfnamefont {H.}~\bibnamefont {K{\"o}ppel}},\
  }\href@noop {} {\emph {\bibinfo {title} {Conical Intersections: Electronic
  Structure, Dynamics \& Spectroscopy}}},\ Vol.~\bibinfo {volume} {15}\
  (\bibinfo  {publisher} {World Scientific Publishing Company Incorporated},\
  \bibinfo {year} {2004})\BibitemShut {NoStop}%
\bibitem [{\citenamefont {Gidopoulos}\ and\ \citenamefont
  {Gross}(ress)}]{Gidopoulos2005}%
  \BibitemOpen
  \bibfield  {author} {\bibinfo {author} {\bibfnamefont {N.~I.}\ \bibnamefont
  {Gidopoulos}}\ and\ \bibinfo {author} {\bibfnamefont {E.~K.~U.}\ \bibnamefont
  {Gross}},\ }\href@noop {} {\bibfield  {journal} {\bibinfo  {journal} {arXiv
  preprint cond-mat/0502433, Phil. Trans. R. Soc. A}\ } (\bibinfo {year}
  {\noop{3001}2014, in press})}\BibitemShut {NoStop}%
\bibitem [{\citenamefont {Hunter}(1975)}]{Hunter1975}%
  \BibitemOpen
  \bibfield  {author} {\bibinfo {author} {\bibfnamefont {G.}~\bibnamefont
  {Hunter}},\ }\href {\doibase 10.1002/qua.560090205} {\bibfield  {journal}
  {\bibinfo  {journal} {Int. J. Quant. Chem.}\ }\textbf {\bibinfo {volume}
  {9}},\ \bibinfo {pages} {237} (\bibinfo {year} {1975})}\BibitemShut {NoStop}%
\bibitem [{\citenamefont {Abedi}\ \emph {et~al.}(2010)\citenamefont {Abedi},
  \citenamefont {Maitra},\ and\ \citenamefont {Gross}}]{Abedi2010}%
  \BibitemOpen
  \bibfield  {author} {\bibinfo {author} {\bibfnamefont {A.}~\bibnamefont
  {Abedi}}, \bibinfo {author} {\bibfnamefont {N.~T.}\ \bibnamefont {Maitra}}, \
  and\ \bibinfo {author} {\bibfnamefont {E.~K.~U.}\ \bibnamefont {Gross}},\
  }\href {\doibase 10.1103/PhysRevLett.105.123002} {\bibfield  {journal}
  {\bibinfo  {journal} {Phys. Rev. Lett.}\ }\textbf {\bibinfo {volume} {105}},\
  \bibinfo {pages} {123002} (\bibinfo {year} {2010})}\BibitemShut {NoStop}%
\bibitem [{\citenamefont {Abedi}\ \emph {et~al.}(2012)\citenamefont {Abedi},
  \citenamefont {Maitra},\ and\ \citenamefont {Gross}}]{Abedi2012}%
  \BibitemOpen
  \bibfield  {author} {\bibinfo {author} {\bibfnamefont {A.}~\bibnamefont
  {Abedi}}, \bibinfo {author} {\bibfnamefont {N.~T.}\ \bibnamefont {Maitra}}, \
  and\ \bibinfo {author} {\bibfnamefont {E.~K.~U.}\ \bibnamefont {Gross}},\
  }\href@noop {} {\bibfield  {journal} {\bibinfo  {journal} {J. Chem. Phys.}\
  }\textbf {\bibinfo {volume} {137}},\ \bibinfo {pages} {22A530} (\bibinfo
  {year} {2012})}\BibitemShut {NoStop}%
\bibitem [{\citenamefont {Abedi}\ \emph {et~al.}(2013)\citenamefont {Abedi},
  \citenamefont {Agostini}, \citenamefont {Suzuki},\ and\ \citenamefont
  {Gross}}]{Abedi2013}%
  \BibitemOpen
  \bibfield  {author} {\bibinfo {author} {\bibfnamefont {A.}~\bibnamefont
  {Abedi}}, \bibinfo {author} {\bibfnamefont {F.}~\bibnamefont {Agostini}},
  \bibinfo {author} {\bibfnamefont {Y.}~\bibnamefont {Suzuki}}, \ and\ \bibinfo
  {author} {\bibfnamefont {E.~K.~U.}\ \bibnamefont {Gross}},\ }\href {\doibase
  10.1103/PhysRevLett.110.263001} {\bibfield  {journal} {\bibinfo  {journal}
  {Phys. Rev. Lett.}\ }\textbf {\bibinfo {volume} {110}},\ \bibinfo {pages}
  {263001} (\bibinfo {year} {2013})}\BibitemShut {NoStop}%
\bibitem [{not()}]{note1}%
  \BibitemOpen
  \href@noop {} {}\bibinfo {note} {We use a numerical grid method with grid
  spacing 0.12 a.u. and 0.3 a.u. for the nuclear and electronic space,
  respectively. The number of grid points are 101 and 81 for each axis in
  nuclear and electronic space, respectively.}\BibitemShut {Stop}%
\end{thebibliography}%

\end{document}